\pdfoutput=0 

\documentclass[epj]{webofc}
\usepackage[varg]{txfonts}   
\usepackage{textcomp} 

\wocname{\includegraphics[width=0.25cm,clip]{logo} ARENA-2016}
\woctitle{ARENA-2016}
\begin{document}
\title{Status of air-shower measurements with sparse radio arrays}
%
%

\author{\firstname{Frank G.} \lastname{Schr\"oder}\inst{1}\fnsep\thanks{\email{frank.schroeder@kit.edu}}
}

\institute{Institut f\"ur Kernphysik, Karlsruhe Institute of Technology (KIT), Germany
          }

\abstract{
This proceeding gives a summary of the current status and open questions of the radio technique for cosmic-ray air showers, assuming that the reader is already familiar with the principles. 
It includes recent results of selected experiments not present at this conference, e.g., LOPES and TREND.
Current radio arrays like AERA or Tunka-Rex have demonstrated that areas of several km$^2$ can be instrumented for reasonable costs with antenna spacings of the order of $200\,$m. 
For the energy of the primary particle such sparse antenna arrays can already compete in absolute accuracy with other precise techniques, like the detection of air-fluorescence or air-Cherenkov light. 
With further improvements in the antenna calibration, the radio detection might become even more accurate. 
For the atmospheric depth of the shower maximum, $X_\mathrm{max}$, currently only the dense array LOFAR features a precision similar to the fluorescence technique, but analysis methods for the radio measurement of $X_\mathrm{max}$ are still under development.
Moreover, the combination of radio and muon measurements is expected to increase the accuracy of the mass composition, and this around-the-clock recording is not limited to clear nights as are the light-detection methods. 
Consequently, radio antennas will be a valuable add-on for any air shower array targeting the energy range above $100\,$PeV.
}
\maketitle
\section{Introduction}
\label{sec_intro}
Immense progress has been made in the last years regarding the understanding of the radio emission by air showers and regarding the development of experimental techniques. 
Because of this progress current radio arrays can measure air-shower observables such as the arrival direction, the energy, and the depth of the shower maximum, $X_\mathrm{max}$, with accuracies similar to those of the established and leading optical techniques. 
While these classical techniques of air-Cherenkov and fluorescence light detection are limited to clear nights, radio detection is possible around the clock.
Only a few percent of the total time is not usable for cosmic-ray physics, since thunderstorm clouds alter the radio signal \cite{2011ApelLOPES_Thunderstorm, SchellartLOFARthunderstorm2015}.
The advantages and the recent success of the radio technique are outlined in detail in longer review articles \cite{HuegeReview2016, SchroederReview2016}, which also contain plenty of introduction to the field. 
This proceeding gives a short summary and points out some open tasks for the future development of the radio technique for air showers.
On the one hand, technical improvements are necessary for using antennas arrays as stand-alone detectors, e.g., the demonstration of an efficient and pure self-trigger.
On the other hand, there are aspects in which the radio technique brings intrinsic advantages over other techniques, e.g., for the detection of inclined air showers ($\theta \approx 70^\circ$) \cite{Gousset2004, AERAinclined_ARENA2016}, or for determination of the cosmic-ray energy scale \cite{AERAenergyPRL2015}.


\section{Understanding of the radio emission}
There is consensus among theorists working in the field, that the radio emission of air-showers is mostly due to the geomagnetic deflection of electrons and positrons \cite{Allan1971, FalckeNature2005, Ardouin2009}, and to a smaller extent by the Askaryan effect due to the time-variation of the negative charge excess \cite{AugerAERApolarization2014, SchellartLOFARpolarization2014, CODALEMAchargeExcess2015, KostuninTheory2015}. 
The strength of the Askaryan relative to the geomagnetic effect for a given geomagnetic angle $\alpha$ depends on the distance to the shower axis and on the zenith angle $\theta$ (see Refs.~\cite{SchroederReview2016, KostuninPhDThesis2015} for a compilation of various measurements). 
There likely is a small phase shift between both emission mechanisms, and first experimental indications by LOFAR and SLAC T-510 have been discussed at this conference. 
Moreover, there is agreement and experimental evidence \cite{CROME_PRL2014, NellesLOFAR_CherenkovRing2014} that the refractive index of the air affects the coherence conditions. 
Therefore, the emission is enhanced at the Cherenkov angle which is around $1^\circ$ for air-showers leading to a Cherenkov ring of about $200\,$m diameter at ground for vertical showers. 
While at lower frequencies $\lesssim 100\,$MHz the signal fills the complete area inside the Cherenkov ring and extends beyond, at higher frequencies of several $100\,$MHz up to a few GHz, the emission is detectable almost exclusively at the Cherenkov angle.

\begin{figure}[t]
 \centering
  ~~
  \includegraphics[width=0.44\textwidth]{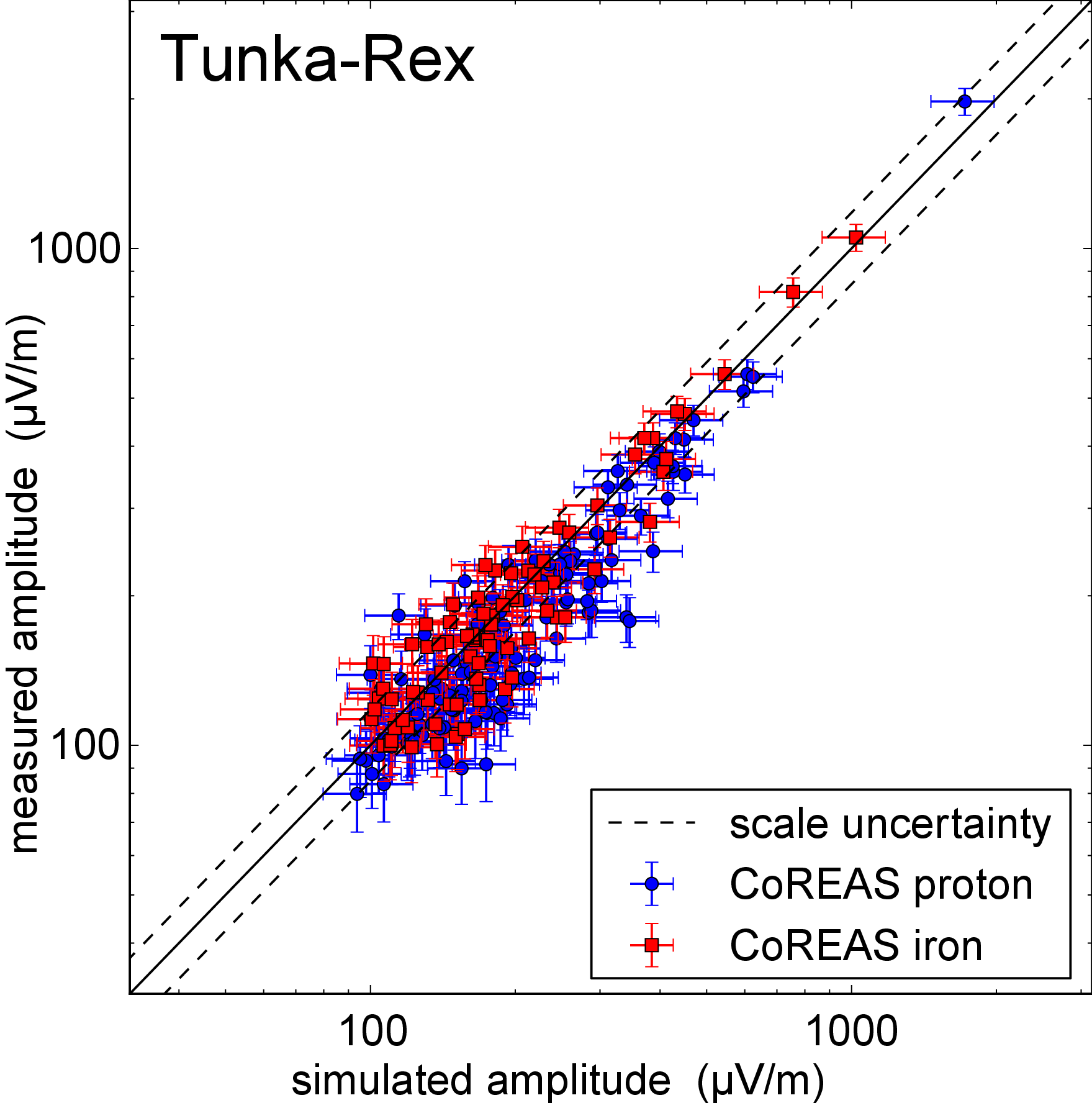}
  \hfill
  \includegraphics[width=0.444\textwidth]{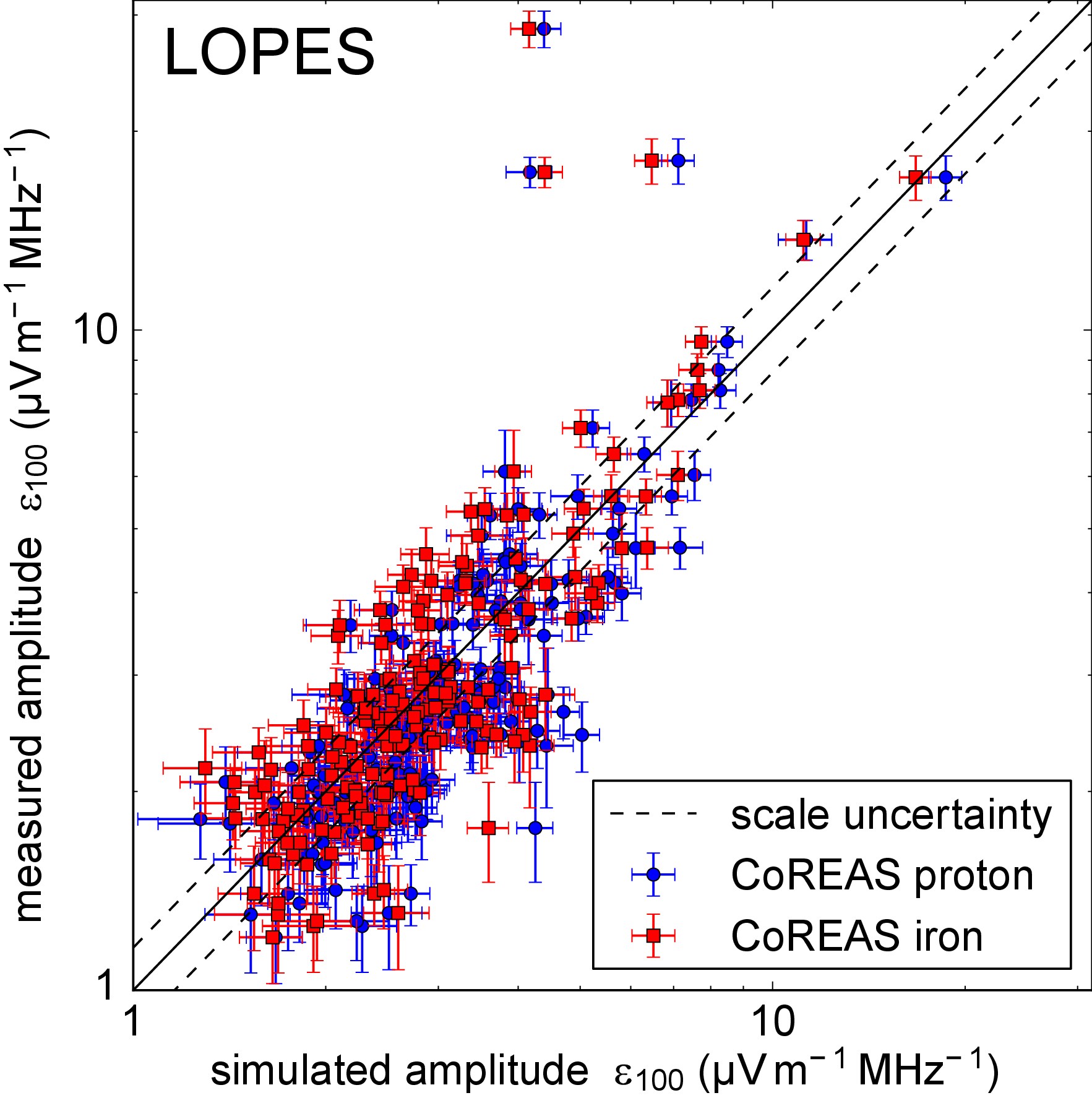}
  ~~
  \caption{Comparison of radio amplitudes measured by Tunka-Rex and LOPES to CoREAS simulations of air-showers initiated by protons and iron nuclei. 
  The energy and shower geometry was set to the values reconstructed by the host experiments Tunka-133 and KASCADE-Grande (from Ref.~\cite{TunkaRexScale2016}). 
  }
  \label{fig_LOPESvsCoREAS}
\end{figure}   

This experimentally confirmed knowledge of the radio emission is implemented implicitly \cite{HuegeCoREAS_ARENA2012, Alvarez_ZHAires_2012} or explicitly \cite{SELFAS2_ARENA2012, Werner_EVA2012} in a variety of codes calculating the radio emission of simulated air showers.
Until now only CoREAS has been extensively tested against measurements: absolute amplitude measurements by LOPES and Tunka-Rex are consistent inside of a $20\,\%$ scale uncertainty with CoREAS (see figure \ref{fig_LOPESvsCoREAS}). 
The apparent disagreement of LOPES and CoREAS published earlier \cite{2013ApelLOPESlateralComparison}, has recently been solved by a reevaluation of the absolute calibration of LOPES \cite{2015ApelLOPES_improvedCalibration}. 
Moreover, Tunka-Rex \cite{TunkaRex_NIM_2015} and LOFAR have been \cite{NellesLOFAR_calibration2015} calibrated with exactly the same reference source as LOPES: 
LOPES and Tunka-Rex measurements agree within the calibration uncertainty \cite{TunkaRexScale2016}, and a comparison with LOFAR measurements has still to be done. 
AERA features an independent absolute calibration \cite{AERAantennaPaper2012}, and the measured radiation energy is at least on average compatible with CoREAS and ZHAireS \cite{AERAenergyPRL2015}. 
Finally, measurements at SLAC under controlled laboratory conditions agree with simulations, but still feature a larger systematic uncertainty of $40\,\%$ \cite{SLAC_T510_PRL2016}.

In summary, the physics of the radio emission seems to be understood to a level of at least $20\,\%$, and a better test is hampered by the calibration uncertainty of current experiments. 
The same calibration uncertainty also limits the scale accuracy of the shower energy determined from radio measurements. 
Currently the scale accuracy is between $10\,\%$ \cite{AERA_calibration_ARENA2016} and $20\,\%$ \cite{2015ApelLOPES_improvedCalibration} for different experiments, where the difference primarily originates from the scale accuracies of the calibration sources claimed by the manufacturers. 
The calibration uncertainty also limits the scale accuracy of the shower energy determined from radio measurements. 
Consequently, one of the most important tasks is to improve the absolute calibrations of antennas. 
Then, the radio technique has a chance to provide an even more accurate measure of the absolute shower energy than the currently leading fluorescence technique, since the strength of the radio signal amplitude depends less on atmospheric conditions.

\begin{table}[t]

\caption{Selected antenna arrays dedicated to air-shower detection, and the local strength of the geomagnetic field according to the international geomagnetic reference model {IGRF} \cite{PotsdamBfield2016}.
Some experiments have been operated in additional configurations, and some are missing in the table, e.g., those located at Antarctica.} \label{tab_experimentsSites}
\small
\begin{tabular}{lcccrrccc}
\hline
Name of & Operation & Latitude & Longitude & $B_\mathrm{geo}$ & Number of & Area & Band\\
experiment & period &  &  & in \textmu T & antennas & in km\textsuperscript{2} & in MHz \\
\hline
Yakutsk \cite{YakutskICRC2015}& since 1972 & $61^\circ 42$' N & $129^\circ 24$' E & $59.7$ & $6$ & $0.1$ & $32$\\
LOPES \cite{SchroederLOPESsummaryECRS2014}& 2003$-$2013 & $49^\circ06$' N & $8^\circ26$' E & $48.4$ & $30$ & $0.04$ & $40-80$\\
CODALEMA \cite{CODALEMAchargeExcess2015}& since 2003 & $47^\circ 23$' N & $2^\circ 12$' E & $47.6$ & $60$ & $1$ & $2-200$\\
TREND \cite{TREND2011}& 2009$-$2014 & $42^\circ 56$' N & $86^\circ 41$' E & $56.3$ & $50$ & $1.2$ & $50-100$\\
AERA \cite{AERAairplanePaper2015}& since 2010 & $35^\circ06$' S & $69^\circ30$' W & $24.0$ & $153$ & $17$ & $30-80$\\
LOFAR \cite{LOFARNature2016}& since 2011 & $52^\circ55$' N & $6^\circ52$' E & $49.3$ &several $1000\,^{*}$ & huge$^{*}$ & $10-240$\\
Tunka-Rex \cite{TunkaRex_NIM_2015}& since 2012 & $51^\circ49$' N & $103^\circ04$' E & $60.4$ & $63$ & $1$ & $30-80$\\ 
SKA-low \cite{HuegeSKA_ICRC2015}& planned & $26^\circ 41$' S & $116^\circ 38$' E & $55.5$ & $60,000$ & $1$ & $50-350$\\
\hline
\multicolumn{9}{l}\footnotesize{$^{*}$ Air showers are measured by several 100 antennas within a few $100\,$m from the center of LOFAR.~~~~~~~~~~~~~}
\end{tabular}
\end{table}

\section{Experiments}
Already in the 1960s a variety of analog radio experiments measured cosmic-ray air showers~\cite{Allan1971}.
In the 2000s digital radio experiments started to successfully measure air-showers, and several second-generation digital arrays do so now. 
Apart from antenna arrays dedicated to this purpose (see table \ref{tab_experimentsSites}), also radio experiments aiming mainly at neutrino searches, such as ANITA \cite{ANITA_CR_PRL_2010} and ARIANNA \cite{ARIANNA_ARENA2016}, have detected cosmic-ray air showers. 
Articles on most of these experiments can be found in this issue, with a few exceptions:
\textbf{LOPES} was the radio extension of the KASCADE-Grande air-shower array at the Karlsruhe Institute of Technology. 
Several methods have been developed by LOPES, e.g., for amplitude and time calibration \cite{NehlsHakenjosArts2007, SchroederTimeCalibration2010}, or for the measurement of the energy \cite{2014ApelLOPES_MassComposition}, and position of the shower maximum \cite{2012ApelLOPES_MTD, 2014ApelLOPES_MassComposition, 2014ApelLOPES_wavefront}.
These methods are now applied to newer experiments in more radio-quiet environments.
While LOPES was stopped in 2013, its data analysis still continues, and the data are planned to be released to the public as part of the KCDC project \cite{KCDC_ECRS2014}. 
\textbf{TREND} was a prototype experiment in Tianshan, China. 
It has shown that self-triggering on the radio signal enables the detection of air showers in this radio-quiet environment \cite{TREND2011}. 
Its successor, GRANDproto, will examine the self-triggering efficiency and purity in more detail, especially for inclined showers, since near-horizontal showers are an interesting target for neutrino detection. 
\textbf{TAROGE} is an antenna tower on top of a mountain in Taiwan \cite{TAROGE_ICRC2015}.
The principle is similar to ANITA except that the antennas are not hanging on a balloon \cite{ANITA_CR_PRL_2010}. 
Observing the ocean, TAROGE aims mainly at the detection of near-horizontal air showers before and after reflection on the water.

\begin{figure*}[t]
  \centering
  \includegraphics[width=0.63\linewidth]{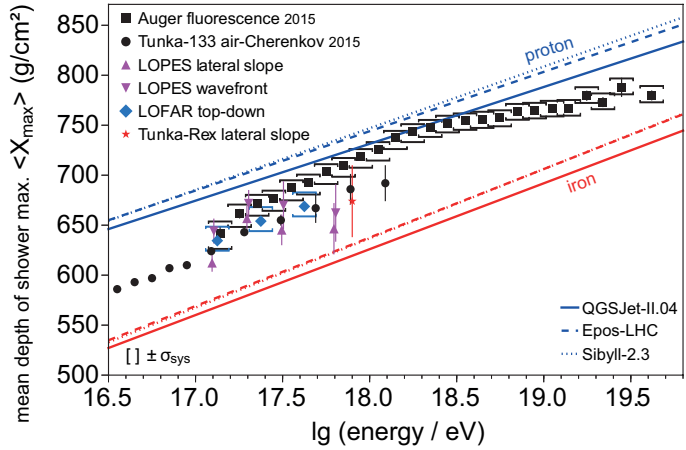}
  \caption{Mean $X_\mathrm{max}$ values measured by radio arrays using different analysis methods, and for reference measurements by the Auger fluorescence detectors and the Tunka-133 air-Cherenkov detectors, and predictions by CORSIKA simulations using various interaction models (from Ref.\cite{SchroederReview2016}, values from Refs.~\cite{Tunka133_ISVHECRI2014, AugerHEATXmaxICRC2015, SchroederLOPESsummaryECRS2014, LOFARNature2016, KostuninPhDThesis2015, RiehnPhDThesis2015}).}
  \label{fig_XmaxOverview}
\end{figure*}

\section{Applications of the radio technique for air showers}

\emph{Ultra-dense arrays}, such as LOFAR and the future SKA \cite{HuegeSKA_ICRC2015}, feature antenna spacings which partly are of the order of a wavelength and can measure the radio signal in great detail. 
Thus, their resolution on the main shower observables (direction, energy, $X_\mathrm{max}$) is not limited by background, but by various systematic uncertainties. 
Including these systematic uncertainties LOFAR already now has a resolution similar to the leading air-fluorescence technique \cite{LOFARNature2016}, but with less exposure than the Pierre Auger Observatory \cite{AugerNIM2015, AugerHEATXmaxICRC2015} or the Telescope Array \cite{TALE_FD_ICRC2015} in the energy range between $10^{17}$ and $10^{18}\,$eV (see figure \ref{fig_XmaxOverview}). 
Although the SKA will be able to compete also in exposure, the real advantage of ultra-dense arrays is that their precision can be even better than that achieved by the optical techniques. 
Nevertheless, methods exploiting mass-sensitive radio observables going beyond a simple analysis of $X_\mathrm{max}$ distributions still have to be developed.
\\
\emph{Radio extensions} of existing air-shower arrays, such as AERA or Tunka-Rex, can enhance the total accuracy by providing an independent measurement of the electromagnetic shower component for relatively little additional costs. 
This use-case of radio extension is the one most advanced, with dedicated software publicly available \cite{RadioOffline2011}.
The co-located particle detector array can provide a fully efficient trigger, which reduces systematic uncertainties due to the dependencies of the radio detection efficiency on the arrival direction and on the time-varying background level. 
It has been shown that even with detector spacings of about $200\,$m, radio measurements of the absolute shower energy are precise and accurate to about $15-20\,\%$ \cite{AugerAERAenergy2015, TunkaRex_Xmax2016}, which is similar to other detection techniques. 
The $X_\mathrm{max}$ precision currently achieved with AERA and Tunka-Rex is of the order of $40\,$g/cm$^2$ \cite{AERA_XmaxMethods_ARENA2016, TunkaRex_Xmax2016}, i.e., twice worse than that of air-fluorescence measurements. 
However, reconstruction methods of $X_\mathrm{max}$ are still under development, and the precision likely can be improved by combining complementary radio observables \cite{AERA_XmaxMethods_ARENA2016}, e.g., the footprint \cite{BuitinkLOFAR_Xmax2014}, the wavefront \cite{2014ApelLOPES_wavefront}, or the frequency spectrum \cite{Grebe_ARENA2012}.
Furthermore, a combined analysis of the radio signal with muon measurements ought to bring additional mass-sensitivity for all zenith angles \cite{Holt_TAUP2015}. 
This is important since it is not yet clear whether the $X_\mathrm{max}$ precision will be sufficient for inclined showers.
\\
\emph{Huge autonomous arrays} larger than the Pierre Auger Observatory, such as the proposed project GRAND \cite{GRAND_ICRC2015}, might provide a way for enhancing the world aperture on the highest energy cosmic rays and additionally for searching neutrinos above $10^{19}\,$eV.
Since the radio footprint extends over several km$^2$ for inclined air showers \cite{AERAinclined_ARENA2016}, detector spacings of $1-2\,$km would allow for reasonable total costs, i.e., below the costs of typical space experiments.
CODALEMA \cite{CODALEMAautonomousRICAP2011}, TREND \cite{TREND2011}, AERA \cite{RAugerSelfTrigger2012}, and ARIANNA \cite{ARIANNA_ARENA2016} have shown that autonomous radio detection is feasible, but there remain some open questions: 
it has to be shown that the trigger can be pure and efficient enough for inclined showers, and there are technical challenges regarding infrastructure, data communication, and reliability of long-term and large-scale remote operation. 
Alternative ways to large apertures might be provided by observing air-showers with antennas in space \cite{MotlochSatelliteExposure2014}, or by using the moon as a target \cite{BrayReview2016}.
However, these methods require successful proof-of-principle demonstrations, and it is not yet clear how they will compare in terms of measurement accuracy to ground-based antenna arrays.

\section{Conclusion}
The original dream of completely replacing air-fluorescence by radio detection might not become true: 
not because the radio technique would not be accurate enough, but simply because the dense antenna spacing required for full-sky coverage makes it more expensive than originally thought. 
Nevertheless, the radio technique likely is a suitable replacement for inclined showers, has a variety of other use cases, and will potentially become even more accurate than the optical techniques.

\section*{Acknowledgements}
Thanks go to the conference organizers, to my colleagues at KIT, to the LOPES, Pierre Auger and Tunka-Rex Collaborations for fruitful discussions, and to DFG for grant Schr 1480/1-1.

\bibliography{arena2016.bib}

\end{document}